\documentclass[aps, prl, reprint, twocolumn, longbibliography]{revtex4-1}  
\usepackage{graphicx,graphics,color,epsfig} 
\usepackage{amsmath, amssymb,amsfonts} 
\usepackage{bm,times,xspace,mhchem} 
\usepackage{ulem} 
\usepackage[colorlinks,citecolor=blue,linkcolor=red]{hyperref} 

\begin{document}

\preprint{AIP/123-QED}

\title{Unsupervised Manifold Clustering of Topological Phononics}

\author{Yang Long}
\author{Jie Ren}
\email{Corresponding Email: Xonics@tongji.edu.cn}
\author{Hong Chen}
\affiliation{%
Center for Phononics and Thermal Energy Science, China-EU Joint Center for Nanophononics, Shanghai Key Laboratory of Special Artificial Microstructure Materials and Technology, School of Physics Sciences and Engineering, Tongji University, Shanghai 200092, China
}%

\date{\today}

\begin{abstract}
Classification of topological phononics is challenging due to the lack of universal topological invariants and the randomness of structure patterns. 
Here, we show the unsupervised manifold learning for clustering topological phononics without any priori knowledge, neither topological invariants nor supervised trainings, even when systems are imperfect or disordered.
This is achieved by exploiting the real-space projection operator about finite phononic lattices to describe the correlation between oscillators.
%The projection operators about phononic systems are exploited for describing the topological responses and correlations between oscillators. 
We exemplify the efficient unsupervised manifold clustering in typical phononic systems, including one-dimensional Su-Schrieffer-Heeger-type phononic chain with random couplings, amorphous phononic topological insulators, higher-order phononic topological states and non-Hermitian phononic chain with random dissipations. %We find that the efficient clustering associated with topological properties can be achieved unsupervisedly. 
The results would inspire more efforts on applications of unsupervised machine learning for topological phononic devices and beyond. 
\end{abstract}
                    
\maketitle

Topological phononics unveil complex mechanism behind unconventional phononic wave phenomena~\cite{liu2017berry, huber2016topological, zhang2010topological, ma2016acoustic, ge2017breaking, xue2019acoustic,zhang2019second}, which lead to the backscattering immune phononic transport modes~\cite{hasan2010colloquium, xiao2010berry, bansil2016colloquium, susstrunk2015observation, wang2015topological}, and would be a promising route to the future robust on-chip communication devices~\cite{huber2016topological, yan2018chip, xiao2015synthetic, serra2018observation, PhysRevLett.119.255901, mousavi2015topologically}. 
%Various topological phenomena have been reported based on their non-trivial topological transitions, including the topologically induced interface modes~\cite{xiao2010berry, xiao2015geometric}, the robust boundary states~\cite{hasan2010colloquium, lu2017observation}, and the pseudo-spin based phononic wave router~\cite{he2016acoustic}. 
To characterize non-trivial phononic topological properties, the key fundamental physical concept is the topological invariant, which is responsible for classifying the topological classes. 
However, there is NO universal topological invariant for all topological phononic systems. And it is even difficult to properly define them when considering many aspects such as symmetry conditions~\cite{mousavi2015topologically}, geometry features~\cite{susstrunk2015observation} and material dispersive responses~\cite{wang2015topological}.
Yet, no matter whether topological invariants are well defined or not, topological properties of phononic states are essentially embedded in the global structure features.
So far, most of topological invariants are defined based on the Bloch momentum space of the perfect periodic structure~\cite{ge2017breaking, he2016acoustic, serra2018observation, he2018topological}. 
But, these momentum-based approaches will have inevitable shortages or inapplicability when handling with phononic models like the spatial randomness of mechanical parameters, the non-Hermitian features or the amorphous structures. 
%The real space descriptions of topological properties that are more generalized than momentum space is exploited instead, which can work for finite systems and is feasible for experimental detections. 
Therefore, finding a general way to explore the topological properties based on real space without defining topological invariants \textit{ad hoc} will be significant but also challenging for the future development of topological phononics and beyond.

Machine learning has shown the power on condensed matters, quantum domains and topological physics~\cite{zhang2018machine, venderley2018machine, peurifoy2018nanophotonic, MEHTA20191,  dunjko2018machine}, e.g., the phase learning of quantum many-body systems~\cite{RevModPhys.91.045002, carrasquilla2017machine, zhang2017quantum, van2017learning, carleo2017solving}, the inverse design of topological optics~\cite{pilozzi2018machine, long2019inverse}, and the optimization of meta-material devices~\cite{liu2018training, ma2018deep, liu2018generative}. However, most of these research works focus on the supervised learning, which can not capture the sample features without a priori knowledge and needs extensive samples with well-defined labels. Recently, the unsupervised learning, which can find potential principles behind raw datasets without labels, has been attracting much attention about its ability on phase detections and classifications in spin systems~\cite{PhysRevB.94.195105, PhysRevE.96.022140, rodriguez2019identifying}, particle explorations in high energy physics~\cite{andreassen2019junipr, baldi2014searching} and efficient material discoveries~\cite{tshitoyan2019unsupervised}. Therefore, unsupervised machine learning would be the meaningful and powerful way to detect and classify topological properties from abundant phononic structures without any priori knowledge about topological mechanism. 

In this Letter,  we demonstrate the unsupervised manifold clustering of topological phononics based on similarities of dynamic properties in real space. The real-space dynamic properties of phononic system are represented by its projection operator $\hat{P}$, which reflects the responses and correlations between oscillators and thus contain the necessary information about the topological properties~\cite{marzari2012maximally, soluyanov2011wannier, bianco2011mapping, kitaev2006anyons}. We firstly show manifold learning can unsuperivsedly learn the features about the finite one-dimensional (1D) Su-Schrieffer-Heeger(SSH) chain efficiently and classify the topological classes  due to nonlinear dimensional reduction~\cite{tenenbaum2000global,roweis2000nonlinear}. Then, based on the real space descriptions,  we successfully demonstrate the unsupervised clustering of several topological phononics cases: (1) disordered phononic SSH chain with random couplings;  (2) Amorphous phononic systems with non-zero local Chern number; (3) Higher-order phononic models. (4) Non-Hermitian phononic chain with random dissipation terms. 
These phononic systems are mapped into points of the manifold space with reduced dimensions based on real space features, and are thus conveniently classified into different groups associated with different topological properties. 
%Our work would pave a way about applications of machine learning on the topological phononics and advance the understandings of topological classifications of novel phononic systems.

%\begin{figure*}[tp!]
%\centering
%\includegraphics[width=\linewidth]{fig_random}
%\caption{The unsupervised clustering of 1D phononics chain with random coupling strength and 2D honeycomb phononics lattice with random mass bias.  } 
%\label{fig:random}
%\end{figure*}

Let us introduce the real-space descriptors of phononic systems. The dynamic equations of phononic system can be written as $\hat{H} |\psi_l \rangle = \Omega_l |\psi_l \rangle$, where $\psi_l$ means the $l$-th mode eigenstate of $\hat{H}$ with the frequency $\omega_l$ ($\Omega_l= \omega_l^2$), $l=1, ..., L$ with $L$ the system size. The time-reversal counterpart will be $\hat{H}^ \dagger|\varphi_l\rangle  =\Omega_l^* |\varphi_l\rangle$. Therefore, the projection operator $\hat{P}$ of cut-off frequency $\omega_c$ is introduced,  as~\cite{mitchell2018amorphous, agarwala2017topological, marzari2012maximally, soluyanov2011wannier, bianco2011mapping, kitaev2006anyons}:
\begin{equation}
\hat{P} = \sum_{\omega_l \leq \omega_c}  |\psi_l\rangle\langle \varphi_l|.
\label{eq:P}
\end{equation}
When the system is Hermitian $\hat{H} = \hat{H}^\dagger$, $\langle \varphi_l| = \langle \psi_l|$. $\hat{P}$ describes the responses and correlations between the phononic oscillators, $P_{ij} = \langle x_j | \hat{P} | x_i \rangle$, playing the role of Green's function. When considering the infinite system with periodic Bloch boundary conditions, the $\hat P$  for $m$-th band can be represented as the sum of Bloch wave functions $\tilde{\psi}_{m\bm{k}}$: $\hat P=\frac{V}{(2\pi)^3}\int_{\rm BZ}d\bm{k} |\tilde{\psi}_{m\bm{k}}\rangle\langle\tilde{\psi}_{m\bm{k}}|$, $V$ is the volume~\cite{marzari2012maximally}. We introduce Gaussian kernel with controlled variance $\varepsilon$ to define the similarity between samples $n$ and $n'$ : 
\begin{equation}
K_{\varepsilon}(n, {n'}) = \exp\left( - \frac{||\hat{P}_n - \hat{P}_{n'}||^2}{2\varepsilon L^2}\right),
\label{eq:similarity}
\end{equation}
where $\hat{P}_{n}$ is a $L\times L$ matrix ($n\in N$), denoting the $\hat{P}$ of the $n$-th sample in a set of $N$ different realizations of the Hamiltonian parameters.  
%Let $\{\hat{P}_n\}$ denote the a set of $N$ samples in $\mathbb{C}^{L\times L}$, where $\hat{P}_n$ is the $L\times L$ matrix, $n\in N$.  
$||\cdot||$ is Taxicab $\mathbb{L}^1$-norm distance, $||A||=\sum_i \sum_j |A_{ij}|$. We can see that Eq.~(\ref{eq:similarity}) reflects the similarity by calculating the projection operator between two phononic samples $n$ and $n'$ such that $K_{\varepsilon}(n, {n'}) \approx 1$ when $\hat{P}_n$ can transform into $\hat{P}_{n'}$ with small deformation. If considering the topological invariant $v$ that is function of $\hat{P}$ and $v(\hat{P}+\Delta \hat{P}) \approx v(\hat{P}) + \frac{\partial v(\hat{P})}{\partial \hat{P}} \Delta \hat{P}$, the difference between different $\hat{P}$ will be responsible for classifying the topological properties $v(\hat{P})$, namely $ |v(\hat{P}_n) - v(\hat{P}_{n'})| \propto ||\hat{P}_n - \hat{P}_{n'}||$. Different topological classes with distinct invariants will have small similarity $K_{\varepsilon}$.
%For 2D cases, the Chern number can be described by the $P$ as $C = \frac{1}{2\pi}{\rm Im}[\log(U V U^\dagger V^\dagger)]$, where $U = P e^{i 2\pi \hat{x}/l_x} P$, $V = P e^{i 2\pi \hat{y}/l_y} P$, $l_x$ and $l_y$ is the scale of system~\cite{agarwala2017topological}.

We adopt the typical manifold learning algorithm: diffusion map~\cite{coifman2006diffusion, tenenbaum2000global,roweis2000nonlinear,coifman2005geometric}, which has been successfully applied for the phase detection and classification of quantum spin systems~\cite{rodriguez2019identifying} based on probabilistic transition processes.
The diffusion process is defined by the local probability transition matrix $T_{n, n'} = \frac{1}{Z_n} K_{\varepsilon}(n, {n'})$, where $Z_n = \sum_{n'=1}^{N}K_{\varepsilon}(n, {n'})$ is the normalization term guaranteeing the probability conservation $\sum_{n'} T_{n, n'} = 1$. 
The global diffusion distance between sample $n$ and $n'$ after $t$ steps can be described by $d_t(n,n') \equiv \sum_{n''}\frac{1}{Z_{n''}} |T^t_{n,n''} - T^t_{n',n''}|^2 = \sum_{j=1}^{N-1} \lambda_j^{2t} |(\phi_j)_n-(\phi_j)_{n'}|^2\geq 0$, where the $\phi_j$ are the $j$-th right eigenvectors of $\hat T$, $\hat T \phi_j = \lambda_j \phi_j$, with the ordered eigenvalues $\lambda_{N-1} \leq \cdots \leq \lambda_2 \leq \lambda_{1} \leq 1=\lambda_0$.
The $j=0$ diffusion mode does not contribute since it is a constant vector. 
It is clear that after long time diffusion $t \rightarrow \infty$, the first few components $\phi_j$ with largest eigenvalues $\lambda_j$ will dominate, which means that we only need a few components $\phi_j$ to represent the samples well so as to reduce the dimension. In particular, the number of the top-ranked largest eigenvalues $\lambda_j$ could determine the number of topological clusters without knowing a priori knowledge~\cite{coifman2005geometric,coifman2006diffusion}.

\begin{figure}[tp!]
\centering
\includegraphics[width=\linewidth]{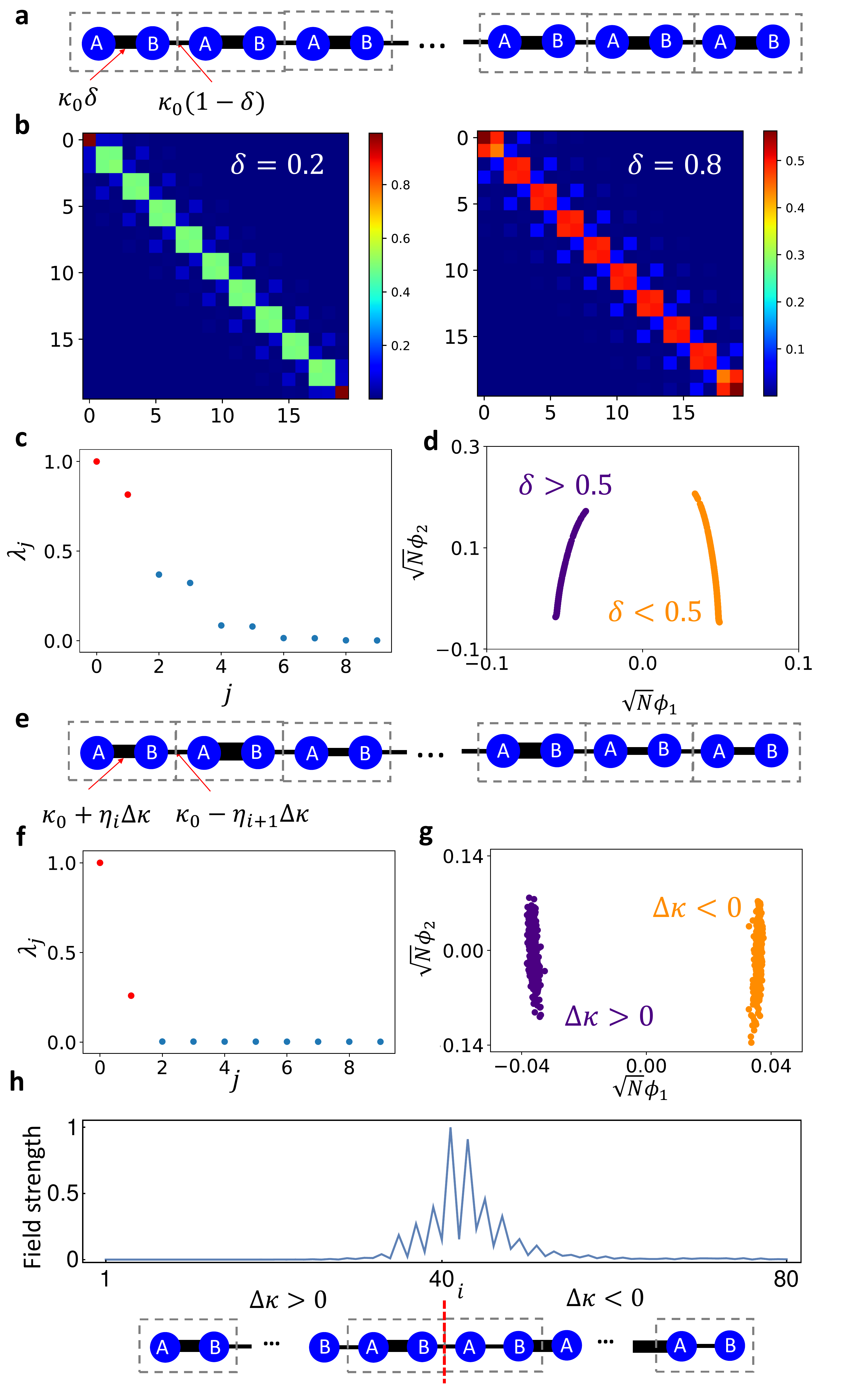}
\vspace{-0.8cm}
\caption{The unsupervised clustering of 1D phononic SSH model with projection operator $\hat{P}$. 
\textbf{a}, the SSH chain with finite size $L=20$, mass $m=1.0$, spring constant $\kappa_0= 1.0$, and $\delta \in (0,1)$. The gray blocks indicate the unit cells. 
\textbf{b}, the projection operator matrix $P_{ij}$ for the cases $\delta=0.2$ and $\delta = 0.8$. 
\textbf{c}, the first 10 largest eigenvalues $\lambda_j$ with $\varepsilon = 0.2$, $N=1000$, $\omega_c = \sqrt{\kappa_0/m} = 1.0$. 
%It is clear that the samples can be obviously classified by the eigenvector $\phi_1$. 
\textbf{d}, different topological classes are classified unsupervisedly, %according to the value $\delta$, 
which coincides with the topological transition ($\delta = 0.5$) in phononic SSH model. 
\textbf{e}, disordered phononic SSH chain with random elastic constants, %It has the length size $L=20$, the mass $m=1.0$, the elastic constant $\kappa_0=1.0$, 
where the bias $\Delta\kappa \in (-0.75, 0.75)$ and the spatially dependent random number $\eta_i \in (0,1)$. 
\textbf{f}, the first 10 largest eigenvalues $\lambda_j$ with $\varepsilon = 1.0$, $N=1000$, $\omega_c=1.4$. 
%It is clear that the samples can be obviously classified by the eigenvector $\phi_1$. 
\textbf{g}, different topological classes can still be well classified, even the systems are disordered by noisy couplings.
%according to the value $\Delta\kappa$, which resembles the topological transition in SSH model.  
\textbf{h}, the topological interface mode emerges between two random phononic chains with opposite signs of $\Delta \kappa$ ($\Delta \kappa = 0.72$ and $\Delta \kappa = -0.54$), and $L=40$. The field strength means the absolute amplitude of displacements. (More details in Supplement.)}
\label{fig:projector}
\end{figure}

\textbf{Topological phononics with random couplings.} We use the  finite 1D SSH phononic chain model~\cite{huber2016topological} to demonstrate the unsupervised clustering of  topological phononics in Fig.~\ref{fig:projector}(a). The dynamic equation can be written as:
\begin{equation}
\begin{aligned}
m \frac{\partial^2}{\partial t^2} a_i &= -\kappa_1 (a_i - b_{i-1}) + \kappa_2 (b_i-a_i), \\
m \frac{\partial^2}{\partial t^2} b_i &= -\kappa_2(b_i - a_i) + \kappa_1 (a_{i+1} - b_i),
\end{aligned}
\end{equation}
where $a_i$ and $b_i$ are the displacements of two atoms in the $i$-th unit cell from their equilibrium positions, the elastic constants $\kappa_1 = \kappa_0 \delta$, $\kappa_2 = \kappa_0 (1-\delta)$, the $\kappa_0$ is a constant spring constant and the random number $\delta \in (0,1)$.  
The calculated projection operator $\hat{P}$ for the different $\delta$ will have some different features as shown in Fig.~\ref{fig:projector}(b), which can be captured and learnt unsupervisedly by the manifold learning.  
Following the scheme described above on the dataset $\{\hat{P}_n\}$, we can see that there is a second high value ($j=1$) in Fig.~\ref{fig:projector}(c), which means that the connections of samples can be reflected by the $\phi_1$. If we assign all the samples in the manifold space $\phi_1$ and $\phi_2$ in Fig.~\ref{fig:projector}(d),  we can see obviously the samples could be classified into two different groups according to the $\delta$, with the threshold value $\delta = 0.5$, coinciding with the topological transition of SSH model. For SSH chain model, the winding number of this system can be calculated by~\cite{susstrunk2016classification}: $v = \frac{1}{2\pi i} \int_{-\pi/a}^{\pi/a}dk q^{-1} \partial_k q $, where $q = \kappa_1 + \kappa_2 e^{i k a} $, which  
%. Based on the residue theorem, the winding number $v$ 
will be $v=1$ for $\kappa_1 < \kappa_2$ ($\delta < 0.5$), $v=0$ for $\kappa_1 > \kappa_2$ ($\delta > 0.5$).

We introduce random elastic constants in the 1D SSH-like phononic chain, as shown in Fig.~\ref{fig:projector}(e). The calculated eigenvalues $\lambda_j$ and the manifold space $\{\phi_j\}$ in Fig.~\ref{fig:projector}(f,g) show that the samples can be also clustered well by our scheme, which resembles the topological transition in standard SSH phononic model for opposite $\Delta \kappa$. The topologically induced interface states for two disordered phononic chains with opposite signs of $\Delta \kappa$ is clearly shown in Fig.~\ref{fig:projector}(h). The successful unsupervised clustering for disordered valley Hall states in 2D honeycomb phononic lattices with random mass biases can be found in Supplement.

\begin{figure}[tp!]
\centering
\includegraphics[width=\linewidth]{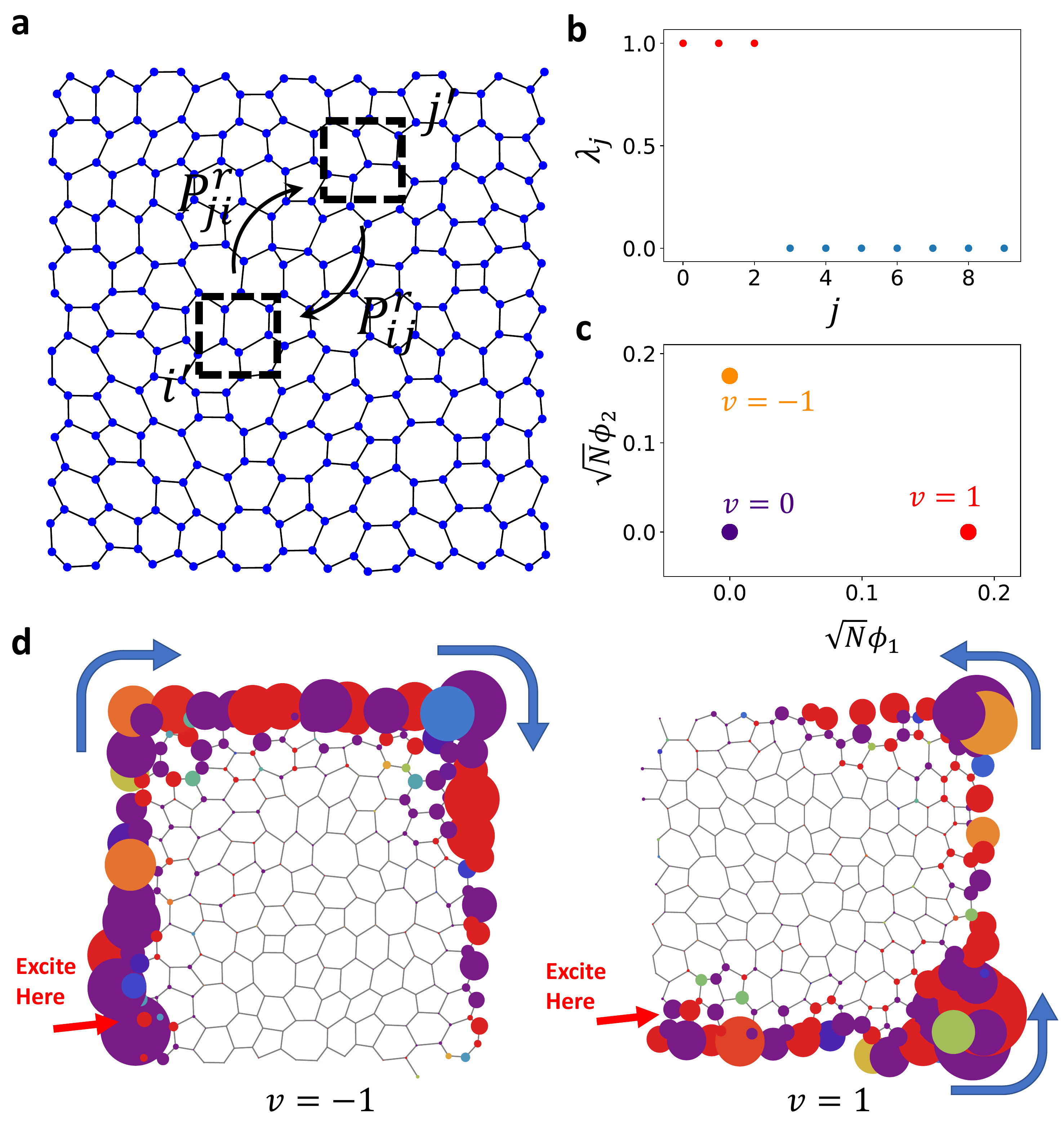}
\vspace{-0.8cm}
\caption{The unsupervised learning of amorphous topological phononics.  \textbf{a}, the demonstration of the amorphous structure, constructed by connecting adjacent centroids of Delaunay triangulation mesh~\cite{florescu2009designer}.  The system has size $L=130\sim150$.  
The projection operator is constructed in the coarse-grained space by averaging displacement field of oscillators in the dotted block of the original space. 
The reduced dimension is $L_r = 10\times10=100$. \textbf{b}, the first 10 largest eigenvalues $\lambda_j$ with $\varepsilon = 0.2$, $N=300$, $\omega_c = 1.0$. \textbf{c}, different classes that are classified unsupervisedly can coincide with different local Chern numbers $v$. The numbers of samples for $v=(0,-1,1)$ are $107$, $93$, $100$, respectively.  \textbf{d}, the chiral edge modes in amorphous topological phononics with different local Chern number $v = \pm 1$. The amplitude and phase of phononic field are represented by the radius and colors (from blue to red) of the disks.}
\label{fig:amorphous}
\end{figure}

\textbf{Amorphous topological phononics insulators.} Recently, the amorphous phononic lattice from random point sets has shown non-trivial topological properties and robust edge states, which demonstrates that the local interactions and local geometry arrangements are sufficient to generate chiral edge modes~\cite{mitchell2018amorphous, agarwala2017topological}, as shown in Fig.~\ref{fig:amorphous}. The amorphous phononic topological systems are constructed by gyroscopes that are linked by springs~\cite{mitchell2018amorphous}, of which the topological properties can be adjusted by the geometric structures and the amorphous types. Because the sample structures are geometrically different due to the randomness, we exploit a coarse-grained mapping from the arbitrary amorphous structures into a uniform space for building uniform descriptions, as shown in Fig.~\ref{fig:amorphous}(a): mapping the average displacement field of oscillators in the dotted block of the real space into the coarse-grained space, and then construct the $\hat{P}^r$ based on the reduced uniform space.

Based on the dataset $\{ \hat{P}_{n}^r\}$, we calculate eigenvalues $\lambda_j$ associated with the manifold space $\{\phi_j\}$ in Fig.~\ref{fig:amorphous}(b,c). We can see the samples can be clustered into three different groups,  coinciding with the toplogical classes of amorphous phononics. For amorphous topological phononics insulators, the local Chern number $v$ is responsible for describing the topological properties~\cite{kitaev2006anyons, mitchell2018amorphous}: $ v = \sum\limits_{i\in A}\sum\limits_{j\in B}\sum\limits_{k\in C} P_{ij}P_{jk}P_{ki}- P_{ik}P_{kj}P_{ji}$, where $A$,$B$ and $C$ is three areas around the positions (See Supplementary). As further demonstrated in Fig.~\ref{fig:amorphous}(d), two different amorphous phononic topological insulators in samples classified by our scheme indeed show opposite chiral edge modes, with opposite local Chern numbers $v=\pm 1$. The samples in the third group do not show any chiral edge mode, corresponding to the trivial topology with $v=0$.

\begin{figure}[tp!]
\centering
\includegraphics[width=\linewidth]{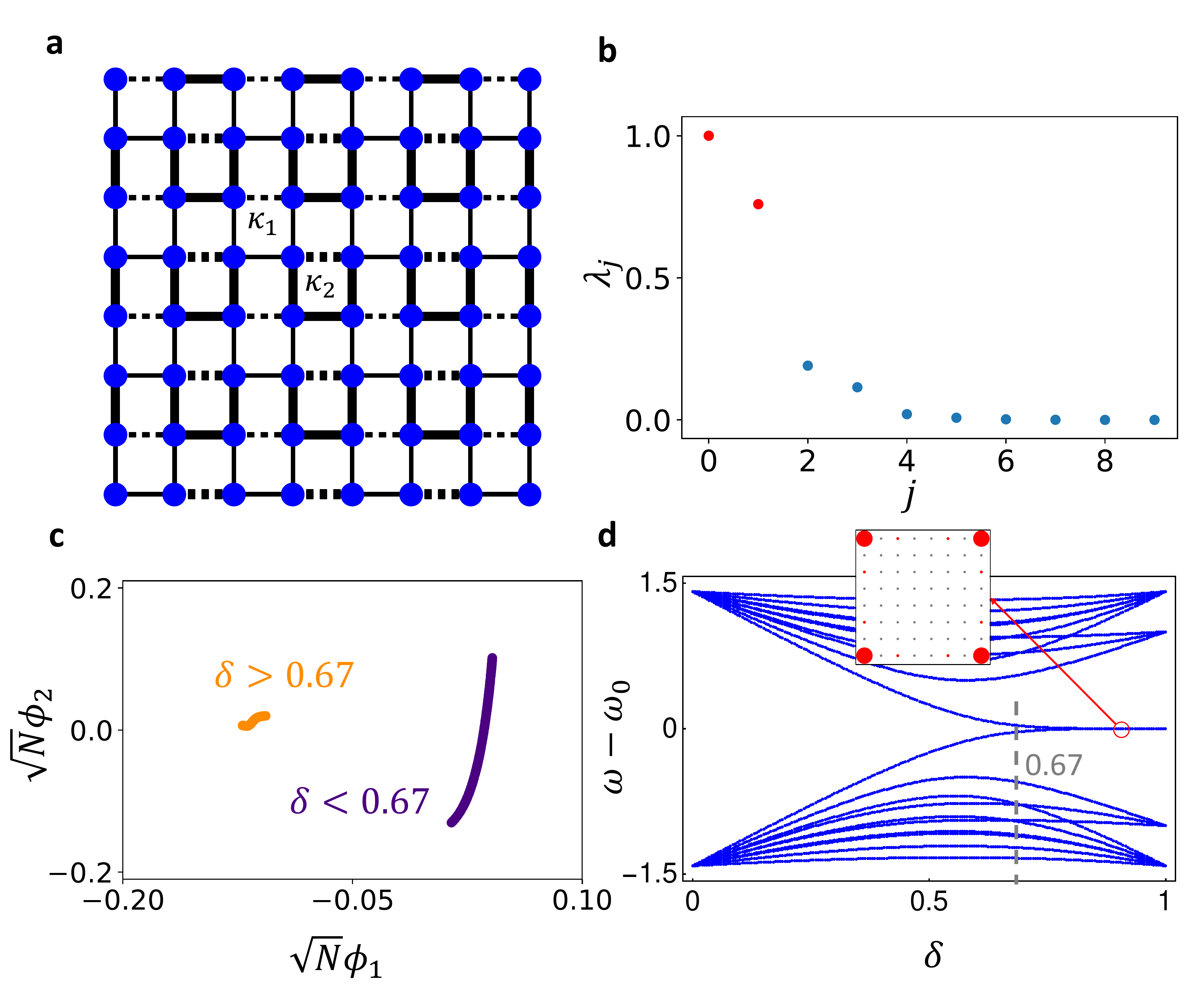}
\vspace{-0.8cm}
\caption{The unsupervised clustering of higher-order topological phononics. \textbf{a}, the quadrupole higher-order topological models with coupling strength $\kappa_1 = \kappa_0 (1- \delta) $ and $\kappa_2 = \kappa_0 \delta$, $\delta \in (0,1)$. The dotted line means the negative counterpart. $L=8\times8=64$. \textbf{b}, the first 10 largest eigenvalues $\lambda_j$ with $\varepsilon = 1.0$, $N= 500$, $\omega_c=\omega_0$. \textbf{c}, the structures are well classified into topological classes, corresponding to different value groups of $\delta$. \textbf{d}, the energy-levels as a function of $\delta$ shows that the unsupervised learning can predict the higher-order topological transition, even at the finite size. Inset shows the quadrupole corner states of the higher-order topological phononics.}
\label{fig:HOTI}
\end{figure}

\textbf{Higher order topological phononics.} Besides these ($d-1$) topological properties in $d$ dimension systems, the higher-order topological states ($d-2$ or $d-3$) are attracting much attention~\cite{benalcazar2017quantized, zhang2019dimensional, zhang2019second, xue2019acoustic, ni2019observation, serra2018observation}. Some higher-order topological phononics can be described by the quantized shift of the Wannier center %$\bm{r}_c = \langle 0, n|\hat{\bm{r}}|n, 0\rangle$, which can be
that is related to the Berry connection~\cite{xue2019acoustic, ni2019observation}. The higher-order topological phononics can be constructed by the continuum phononic systems and the dynamic equations can be approximately written as $\ddot{x}_i = D_{ij} x_j$ around the resonant frequency $\omega_0$~\cite{serra2018observation, matlack2018designing}, where $D_{ij}$ is the effective coupling between the oscillator $x_i$ and the oscillator $x_j$. 

The unsupervised learning of the quadrupole higher-order topological insulator (HOTI) is demonstrated in Fig.~\ref{fig:HOTI}(a). From the eigenvalues $\lambda_j$ associated with the manifold space $\{\phi_j\}$ in Fig.~\ref{fig:HOTI}(b,c), we can see that the $N$ samples can be classified unsupervisedly by the threshold value $\delta = 0.67$. This value deviates from the theoretical one $\delta=0.5$ predicted in Bloch-momentum space analysis~\cite{benalcazar2017quantized,serra2018observation}, due to the finite size effect.  
As shown in Fig.~\ref{fig:HOTI}(d), for the finite system, the topological transition point $\delta$, where the corner states emerge and the frequencies become degenerate at the resonant frequency ($\omega_0$, as the effective zero energy) will shift from theoretical value $0.5$ to $0.67$. 
This is because the ``corner" states, although decay spatially, will interact with each other when the system size is finite. A perfect phase transition requires a clear separation of those corner states, which will be achieved when further increase $\delta$ to $0.67$.
This finite size effect of HOTI classification can not be found before in the Bloch picture for periodic infinite systems. The unsupervised learning for HOTI in phononic Kagome lattices can be found in Supplement.

\begin{figure}[tp!]
\centering
\includegraphics[width=\linewidth]{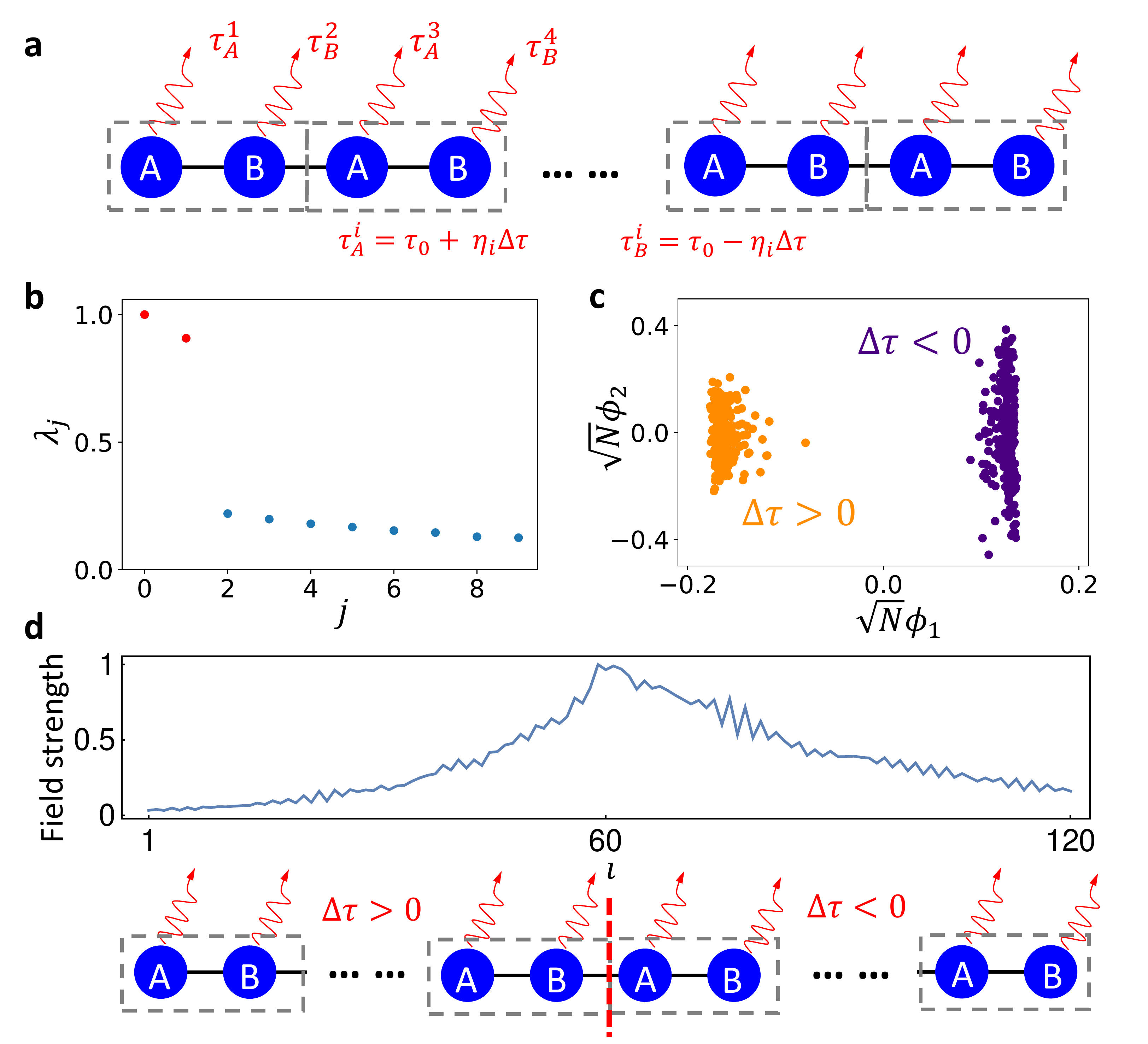}
\vspace{-0.8cm}
\caption{The unsupervised clustering of 1D non-Hermitian phononic chain. \textbf{a}, the A/B oscillators have random dissipation losses, with alternatively large and small dissipation values $\tau_{A/B}^i = \tau_0 \pm \eta_i \Delta \tau$, respectively. The length size $L=20$, the mass $m=1.0$, the viscosity $\tau_0 = 1.0$, the viscosity bias $\Delta \tau \in (-0.5,0.5)$ and the random number $\eta_i \in (0,1)$. \textbf{b}, the first 10 largest eigenvalues $\lambda_j$ with $\varepsilon = 1.0$, $N=500$, $\omega_c = 1.35$. \textbf{c}, different classes are classified unsupervisedly, which coincides with the fact that the cases of $\tau_A > \tau_B$ ($\Delta \tau>0$) and $\tau_A < \tau_B$ ($\Delta \tau<0$) are topologically distinct.  \textbf{d}, the topological interface mode emerges between two non-Hermitian random phononic chains with opposite signs of $\Delta \tau$ ($\Delta \tau = 0.4$ and $\Delta \tau = -0.24$), and $L=60$.}
\label{fig:nonhermitian}
\end{figure}

\textbf{Non-Hermitian topological phononics.} The topological properties in non-Hermitian systems without time-reversal symmetry are very important~\cite{diehl2011topology, ding2016emergence, zeuner2015observation, leykam2017edge,ding2018experimental, shen2018topological,yao2018edge,dasbiswas2018topological,gong2018topological}.  
%So far, the non-Hermitian topological systems involve in many physical properties.
The non-Hermitian feature can be originated from many aspects~\cite{ghatak2019new}, such as the non-reciprocal coupling between the nearest oscillators~\cite{lee2016anomalous},  the non-zero dissipation bias, or the complex on-site potentials~\cite{wang2018valley, leykam2017edge}. Mathematically, the non-Hermitian system will have complex eigenfrequencies and non-orthogonal eigenvectors~\cite{bender2007making}. Some Bloch analysis have been applied for non-Hermitian systems and found that the exception points play important roles for the topological origin~\cite{leykam2017edge, gong2018topological}. However, the non-Bloch analysis also states the non-Hermitian skin (surface) effect and describes the edge modes as well~\cite{shen2018topological,yao2018edge}. Thus, the non-Hermitian topological invariants are not clear in general and still being under explorations~\cite{shen2018topological,liu2019second,yao2018edge,dasbiswas2018topological,gong2018topological}. 

Here, we consider the 1D phononic lattice with different and random dissipation terms due to the non-zero viscosity $\tau_{A/B} \neq 0$, shown in Fig.~\ref{fig:nonhermitian}(a). The dynamic equation can be written as:
\begin{equation}
\begin{aligned}
m \frac{\partial^2}{\partial t^2} a_i + \tau_A^i \frac{\partial}{\partial t}a_i &= - \kappa (a_i - b_{i-1}) + \kappa (b_i - a_i), \\
m \frac{\partial^2}{\partial t^2} b_i + \tau_B^{i+1} \frac{\partial}{\partial t}b_i &= - \kappa (b_i - a_i) + \kappa (a_{i+1} - b_i),
\end{aligned}
\end{equation}
where $\tau_{A/B}^i$ is the different and random dissipation viscosity terms of $A/B$ oscillators for the different $i$-th oscillator. From the eigenvalues $\lambda_j$ associated with the manifold space $\{\phi_j\}$ in Fig.~\ref{fig:nonhermitian}(b,c), we can see the $N$ samples are clustered into two different groups: $\Delta \tau > 0$ or $\Delta \tau < 0$ corresponding to the case $\tau_A>\tau_B$ or $\tau_A < \tau_B$, which coincides with the fact that the cases of $\tau_A > \tau_B$ and $\tau_A < \tau_B$ are topologically distinct. 
As a result, the topological interface state emerges at the interface between two topologically different random non-Hermitian chains with opposite signs of $\Delta\tau$, as shown in Fig.~\ref{fig:nonhermitian}(d).

%Finally, we %discuss the unsupervised clusterings with the realistic experimental data. 
%note that the real space projection based descriptions makes the unsupervised clustering feasible and accessible in experiments and detections.
%Considering that we have collected the evolution trajectory for each oscillator in real space, we can obtain the resonant modes and frequencies by applying the Fourier transformation on the experimental data. 
%One then can construct the $\hat{P}$ based on these mode profiles according to the Eq.~\ref{eq:P} and process the same manifold learning process on the dataset $\{\hat{P}_n\}$. 
%Sometimes there would be some noisy signals contained in the experimental data, we find that that our method will still be robust. (See Supplementary)

To summarize, we have exemplified the function of unsupervised manifold clustering of topological phononics.
Several main difficulties of topological phononics including the spatial randomness, the amorphous non-periodic structures and the non-Hermitian dissipations have been discussed.  
The unsupervised manifold learning have achieved the efficient non-linear dimension reductions, which would map the phononic systems into the manifold space based on their features in the real space and then cluster them into different groups associated with different topological properties.  
Our work would be used to explore diverse topological phononics before defining or introducing topological invariants,  which is meaningful for not only theoretical understandings but also experimental detections, especially for the random, non-Hermitian and out-of-equilibrium open phononic systems~\cite{shen2018topological,yao2018edge,dasbiswas2018topological,gong2018topological}.

\begin{acknowledgments}
We acknowledge the support from the National Key Research Program of China (No. 2016YFA0301101), National Natural Science Foundation of China (No. 11935010, No. 11775159 and No. 61621001), the Shanghai Science and Technology Committee (Nos. 18ZR1442800 and 18JC1410900), the Opening Project of Shanghai Key Laboratory of Special Artificial Microstructure Materials and Technology.
\end{acknowledgments}

%\bibliography{reference}

\end{document}